\documentclass{ws-p9-75x6-50}

\newtheorem{theorem}{Theorem}

\begin{document}

\title{Some results on the integrability of Einstein's field equations for 
axistationary perfect fluids}

\author{M.\ Bradley$^\dag$, G.\ Fodor$^{\ddag,\S}$, M.\ Marklund$^\ast$ 
  and Z.\ Perj\'es$^\S$}

\address{$\dag$ Department of Plasma Physics, Ume{\aa} University, 
  SE--901 87 Ume{\aa}, Sweden \\
  $\ddag$ Department of Physics, Waseda University, 
  3--4--1 Okubo, Shinjuku, Tokyo 169--8555, Japan \\
  $\S$ KFKI Research Insitute for Particle and Nuclear Physics, 
  Budapest 114, P.O.B.\ 49, H--1525 Hungary \\
  $\ast$ Department of Electromagnetics, Chalmers University of 
  Technology, SE--412 96 G\"oteborg, Sweden}

\maketitle

\abstracts{
Using an orthonormal Lorentz frame approach to axistationary 
perfect fluid spacetimes, we have formulated the necessary and sufficient 
equations as a first order system, and investigated the integrability 
conditions of this set of equations. The integrability conditions are 
helpful tools when it comes to check the consequences and/or compatibility
of certain simplifying assumptions, e.g.\ Petrov types. Furthermore, 
using this method, a relation between the fluid shear and vorticity
is found for barotropic fluids. We collect some results concerning Petrov 
types, and it is found that an incompressible axistationary perfect fluid 
must be of Petrov type I.
}

We are interested in axistationary perfect fluid spacetimes, 
i.e.\ the energy-momentum tensor takes the form
$  T_{ab} = (\mu + p)u_au_b - pg_{ab}$,
and the indices $a, b, ...$ refere to a Lorentz frame comoving with the 
fluid (here $p$ is pressure, $\mu$ density). 
Furthermore the spacetime has two commuting Killing vectors 
$\partial/\partial t$ and $\partial/\partial\phi$, and all our 
quantities depend only on the coordinates $x$ and $y$. The frame 
$\{ e_a \}$ is adapted to the spacetime symmetry, i.e., 
$  e_1 = \xi_1\partial_x + \upsilon_1\partial_y$, 
$  e_2 = \xi_2\partial_x + \upsilon_2\partial_y$, 
while $e_0$ and $e_3$ do not contain $\partial_x$ or $\partial_y$. 
With these choices, we bring the Riemann tensor and the Ricci 
rotation coefficients to ``standard form'' for axistationary 
perfect fluids.\cite{Fodor-Marklund-Perjes}
Inherent in this standard form is that we still have 
the freedom to perform a rotation in the tetrad plane spanned by 
$e_1$ and $e_2$, which may be used to reduce the number of variables,
and thus simplify the equations.\cite{Fodor-Marklund-Perjes} 

Our variables -- the Riemann tensor, the Ricci rotation coefficients, 
and the tetrad vector components -- are constrained by (a) the Bianchi 
identities, (b) the Ricci identities, and (c) the commutator equations  
 $ [e_1,e_2] = 2\Gamma^a\!_{[21]}e_a$,  
 $ a = 1, 2$, respectively  
($\Gamma^a\!_{bc}$ being the Ricci rotation coefficients).
These equations form a set equivalent to Einstein's
equations when the Ricci tensor is related to the energy-momentum 
tensor.\cite{Fodor-Marklund-Perjes,Bradley-Karlhede,Bradley-Marklund}

The procedure of checking integrability of this first order system 
is to apply the commutator to equations (a) and (b). We know that
(and this is straightforward to check) the Bianchi identities
are the integrability conditions of the Ricci identities. Thus, in 
order to guarantee integrability, we apply the commutator to the Bianchi
identites. The integrability condition of the momentum conservation 
equations 
$  e_a(p) = \dot{u}_a(\mu + p)$, 
$\dot{u}_a$ being the fluid acceleration -- is
\begin{equation}\label{eq:condition}
  \dot{u}_1e_2(\mu) - \dot{u}_2e_1(\mu) = 
    4(\mu + p)(\omega_1\sigma_2 - \omega_2\sigma_1) \ ,
\end{equation}
$\omega_a$ and $\sigma_a$ being components of the fluid vorticity and 
shear, respectively (for the definitions, see Ref.\ 1). 
It is known that rigid fluids (i.e.\ $\sigma_1 = \sigma_2 = 0$) 
must have a barotropic equation of state,\cite{Senovilla} but using 
(\ref{eq:condition}) we can state a more general result: 
\begin{theorem}
  The equation of state is barotropic if and only if 
  $\omega_1\sigma_2 = \omega_2\sigma_1$.
\end{theorem}

The investigation of the integrability of the system of equations
shows that the four Bianchi identities are satisfied due to the
Ricci identities.
There remain 4 equations (from the Bianchi 
identities) for four components of the Weyl tensor, and the two 
momentum conservation equations 
for the pressure, and these equations
will have solutions according to the Cauchy--Kowalewski theorem.%
\cite{Courant-Hilbert}

Next we list some results concerning some special subcases.

$\bullet$ Perfect fluids with purely magnetic Weyl tensor%
\cite{Fodor-Marklund-Perjes} can be investigated in the special 
cases (a) incompressible fluid, when the Bianchi identities imply 
    conformal flatness, and by a theorem due to Collinson\cite{Collinson}
    this is the interior Schwarzschild solution, and
    (b) rigidly rotating fluid (i.e., shear-free), for which
    all solutions exhibit local rotational symmetry\cite{Ellis} which 
    means that they either are the interior Schwarzschild solution,
    or NUT-like fluids.\cite{Fodor-Marklund-Perjes,Fodor-etal}

$\bullet$ Petrov types can be investigated when our orthonormal 
frame is related to a null tetrad, which will give the Weyl spinor
in terms of the magnetic and electric parts of the Weyl tensor.
(a) Petrov type III is inconsistent with the axistationary 
    assumption,\cite{Fodor-Marklund-Perjes}
(b) Petrov type N turns out to be equivalent to a vacuum
    spacetime with a cosmological constant $\Lambda = -p$;%
    \cite{Fodor-Marklund-Perjes}
(c) using the perturbative approach of Hartle\cite{Hartle}
    combined with the tetrad formalism, Fodor \& Perj\'es%
    \cite{Fodor-Perjes} have shown that
    (i) Petrov type II reduce to the de Sitter spacetime
        in the static limit, and
    (ii) Petrov type D cannot be incompressible;
(d) axistationary spacetimes with purely magnetic or electric
    Weyl tensor must be of Petrov type D.\cite{Fodor-Marklund-Perjes}

As a consequence of the above we can state the following:
{\it A physically realistic rotating incompressible axistationary 
perfect fluid star model must be of Petrov type I}.
Since it can be expected that the simplest physical axistationary
perfect fluid spacetime, that can represent an astrophysical object
with compact support, will be the incompressible ditto (compare with
the interior Schwarzschild spacetime), it is reasonable to assume that 
algebraic restrictions on the Weyl tensor will not be fruitful in 
the search for such solutions.

\end{document}